\documentclass[11pt]{article}
\usepackage{amsmath,amssymb,amsfonts,amsthm}
\usepackage[ruled]{algorithm2e}
\pdfoutput=1

\newtheorem{lemma}{Lemma}[section]

\newtheorem{theorem}{Theorem}[section]

\begin{document}

\setcounter{footnote}{0}
\addtocounter{footnote}{1}
\footnotetext{Department of Applied Mathematics, University of Maryland, 
College Park, MD 20742.
Email: \texttt{davidgharris29@hotmail.com}.}
\addtocounter{footnote}{1}
\footnotetext{Department of Computer Science, University of Maryland, 
College Park, MD 20742.
Email: \texttt{manishp@cs.umd.edu}.}

\author{David Harris$^{1}$, Manish Purohit$^{2}$}
\title{
Improved algorithms and analysis for the laminar matroid secretary problem}

\maketitle

\begin{abstract}
In a matroid secretary problem, one is presented with a sequence of objects of various weights in a random order, and must choose irrevocably to accept or reject each item. There is a further constraint that the set of items selected must form an independent set of an associated matroid. Constant-competitive algorithms (algorithms whose expected solution weight is within a constant factor of the optimal) are known for many types of matroid secretary problems. We examine the laminar matroid and show an algorithm achieving provably $0.053$ competitive ratio.
\end{abstract}

\section{Introduction}
In the classical secretary problem, one interviews $n$ secretaries sequentially in random order, each order having equal probability. As soon as one interviews a secretary, one learns the skill level of that secretary, relative to all previously seen applicants. At this point the interviewer must make an irrevocable decision whether or not to hire. The goal is to hire the best secretary. 

For this problem, \cite{sec-cite1},\cite{sec-cite2},\cite{sec-cite3} discuss the elegant optimal algorithm. This algorithm looks at the first $\frac{n}{e}$ secretaries, rejects them all, and then from among the remaining secretaries chooses the first one who is better than each of the first observed $\frac{n}{e}$ secretaries (if any). This simple algorithm hires the best secretary with probability $\frac{1}{e}$.

One of the many generalizations of the secretary problem is called the matroid secretary problem. Here, we are given a matroid $\mathfrak{M}(\mathcal{U},\mathcal{I})$ (which is known completely beforehand). The ground set also contains weights for each element, which are unknown a priori. The elements arrive one by one in a random order. We denote this ordering by $\pi$, a permutation on $n$ elements. Each element reveals its weight when it arrives. As before we must make an irrevocable decision whether to accept or reject the element when it arrives.  The goal is to choose an independent set of the largest weight. 

A matroid is a particularly attractive setting for the secretary problem, because of the exchange property. This ensures that even if we make a bad decision about which element to accept, we are not locked in to a bad solution set. In matroid secretary problems, as opposed to more general secretary problems, we can often find a solution set which is relatively close to the optimal one.

The matroid secretary problem can be viewed as a simple model for irrevocable decisions in the presence of uncertainty as to future opportunities. The use of random permutation is conceptually simple, but allows powerful bounds with a minimum of auxiliarly information. Other models which may include prior distributions on the price structure are possible.

 For secretary problems, we define the \emph{competitive ratio} to be the ratio of the expected weight obtained by our algorithm, divided by the optimal weight. We note that in the classical secretary problem, one has a $1/e$ chance of choosing the best applicant; for the matroid secretary problem, we do not care about the probability of selecting the largest-weight independent set from the matroid, only in selecting sets which have large weight on average. Furthermore, we do not need any probability of obtaining a large-weight set (other than is implied by Markov's inequality).

For general matroids, \cite{chakraborty} gives an $O(\sqrt{\log r})$-competitive algorithm where $r$ is the rank of the matroid. For many special classes of matroids, constant-competitive algorithms are known. In particular, \cite{laminar} provides the a $\frac{3}{16000}$-competitive algorithm for laminar matroids. An alternative algorithm has been demonstrated in \cite{laminar2}, which gives a $0.070$-competitive algorithm for the laminar matroid.

We improve the algorithm of \cite{laminar} and obtain a tighter analysis, showing a $0.053$-competitive algorithm for the laminar matroid. This improves on \cite{laminar} by nearly $300$-fold. This nearly brings the algorithm of \cite{laminar} to parity with the new algorithm of \cite{laminar2}.

\section{Definitions and Notation}
We let $U$ be the ground set and $w : U \rightarrow \mathbb{R}$ be the weight function. Then a laminar matroid is defined by a family $\mathcal{F}$ of laminar subsets. That is, for any $A, B \in \mathcal F$ we have $A \subseteq B$ or $B \subseteq A$ or $A \cap B = \emptyset$. In other words, the sets in $\mathcal F$ are nested within each other.  Each set $A \in \mathcal{F}$ has an associated \emph{capacity} $\mu(A)$. A set $X \subseteq U$ is an independent set in the matroid iff $|X \cap A| \leq \mu(A)$ for all $A \in \mathcal F$.

Without loss of generality we may assume $\mu(A) < \mu(B)$ for any $A,B \in \mathcal{F}$ and $A \subseteq B$; for, otherwise $A$ is redundant and may be removed from $\mathcal F$. 

We use the terminology of \cite{laminar}. For $i \in U$, we let $M(i)$ denote the minimal set $B \in \mathcal{F}$ such that $i \in B$. We say that $B_1 \in \mathcal{F}$ is a child of $B_2 \in \mathcal{F}$ if $B_1 \subsetneq B_2$ and there exists no intermediate set $B' \in \mathcal{F}$ such that $B_1 \subseteq B' \subseteq B_2$. Naturally $B_2$ is called parent of $B_1$.

For any $A,B \in \mathcal{F}$ such that $A \subseteq B$, we define $\text{Chain}[A,B]$ to be the sequence of sets in $\mathcal{F}$ starting with $A$ and ending with $B$ where each set is a child of the following set. In order to denote all sets in $\mathcal{F}$ that contain $i$, we may interchangeably use $\text{Chain}[M(i), U]$ or $\mathcal{F}(i)$. To save notation, let $\text{OPT}$ denote the optimal solution itself or the total weight of the optimal solution depending on the context. For any $V \subseteq U$ and $B \in F$, let $\text{OPT}_V(B)$ denote the optimal feasible solution that can be obtained from $V \cap B$. For simplicity of notation, let $\text{OPT}(B) = \text{OPT}_U(B)$. Let $\pi$ denote the random ordering of elements in $U$.

\section{Algorithm}

 \ \\
\begin{algorithm}[H]
{\bf Let} Draw $t \sim \text{Binom}(n, 1-p)$ and let $S = \{\pi(1), \dots, \pi(t) \}$.\\
\ForEach{$B \in \mathcal F$} {
{\bf let} $R(B) \leftarrow OPT_S(B)$
}
\ForEach{$i \in T = U - S$ (taken in the random order $\pi$)} {
\ForEach {$B \in \text{Chain}[ M(i), U ]$} {
\eIf{$R(B) \neq \emptyset$ and $w(i)$ is greater than some element of $R(B)$} {
Add $i$ to $SOL(B)$ \\
Remove the largest element of weight less than $w(i)$ from $R(B)$
}
{
break the loop (go to the next item $i$) 
}
}
}
Return $SOL(U)$\;
\caption{KickNext Algorithm}
\label{Alocalsearch}
\end{algorithm}

Here, we take the first $1-p$ proportion of items for the sampling phase (used to estimate statistical information about the optimal solution), and we take the latter fraction $p$ to actually build the optimal solution. As we will see, the optimal choice of $p$ is about $p \approx 0.08$. From the sampled set of elements $S$, we calculate $OPT_S(B)$ as the reference set $R(B)$.

We denote the $S = \{\pi(1), \dots, \pi(t) \}$. Such elements are used for sampling and building statistical information about the optimal set. The remaining items $T = U - S$ are considered for actual selection.

Note that this algorithm does not use the ``AddIt" method used in \cite{laminar}, in which during the second phase items enter the optimal solution with some probability less than one. The intuitive explanation for this difference is that any element which is not eligible for the optimal solution should be used to build statistical information, and not simply discarded. 

We will briefly explain the intuition behind this algorithm. In the initial sampling phase, we build up a set which looks like the globally optimal solution; in the second phase, we try to mimic the sample optimum as closely as possible. The rule for evicting elements from $R(B)$ appears strange, in that it would be more natural to remove the \emph{lowest-weight} element from $R(B)$ when inserting a new element. However, if we did this, then for low-weight elements $R(B)$ would become distorted compared to $OPT(B)$. The key innovation of \cite{laminar} was in using this counter-intuitive eviction rule.
\section{Analysis}

Note that the algorithm selects an element by kicking out a smaller element in $R(B)$ for all $B \in \mathcal{F}(i)$. An element $i$ is not selected to be in $SOL(B)$ iff all elements with weight smaller than $w_i$ in $R(B)$ have been kicked out already. 

We assume that, at the end of the sampling phase, we have $|OPT_S(B)| = \mu(B)$ exactly for all $B \in \mathcal F$. We can force this to occur with probability one by adding infinitely many elements of infinitesimal weight to the matroid, which will not affect the algorithm's behavior. This simplifying assumption allows us to avoid some corner cases. 

Finally, we assume that items have distinct weights; this can be achieved by adding infinitesimal perturbations to the original weights. This affects the behavior of the optimal algorithm only infinitesimally. The perturbation may affect the behavior of this algorithm substantially, as it is based on determining hard cut-off values for whether to accept an element. However, it will suffice to show a good competitive ratio on the perturbed weights.

For a given set $B \in \mathcal F$, most elements $x \in T$ will be immediately disqualified from affecting $B$ in any way. We can note a simple condition on element $x \in B$ affecting the set $\text{SOL}(B)$ is that the weight of $x$ exceeds the smallest weight element of $\text{OPT}_S(B')$, for all $B'$ in the chain between $M(x)$ and $B$. We call such elements \emph{qualifying for $B$}. We can bound the number of such qualifying elements as follows:

\begin{lemma}
\label{Aquallemma}
Consider any set $B \in \mathcal F$ and element $i \in U$. Let $\text{OPT}_S(B) = \{a_1, a_2, \dots, a_m \}$ sorted so that $w(a_1) < w(a_2) < \dots < w(a_m)$. For notational convenience, set $w(a_{m+1}) = \infty$. Let $N_j \subseteq T$, for $j = 1, \dots, m$, denote the elements $x$ which satisfy the following conditions:
\begin{enumerate}
\item $x$ qualifies for $B$
\item $w(a_j) < x < w(a_{j+1})$
\item $x \in T$
\end{enumerate}
Then for any non-negative integers $n_1, \dots, n_m$, we have
$$
P(|N_1| = n_1 \wedge \dots \wedge |N_m| = n_m \mid i \notin S) \leq p^{n_1 + \dots + n_m}
$$
\end{lemma}
\begin{proof}

It suffices to show that, for any $j = 1, \dots, m$, the probability that  $|N_j| = n_j$, conditional on $i \notin S$ as well as $|N_{j+1}| = n_{j+1}, \dots, |N_m| = n_m$, is at most $p^{n_j}$.

Note that $N_j$ is determined solely by the elements of weight less than $w(a_{j+1})$. Suppose we condition on some choice of $a_{j+1}, \dots, a_m$. Now $N_j$ depends solely on the positions of elements with weights less than $w(a_{j+1})$, and in particular is independent of $N_{j+1}, \dots, N_m$. Then $a_j$ is the element of $U$ satisfying the five conditions:
\begin{enumerate}
\item $a_j \neq i$
\item $w(a_j) < w(a_{j+1})$
\item $\{a_j, a_{j+1}, \dots a_m \} \in \mathcal I$
\item $a_j \in S$
\item $a_j$ has maximal weight among all that satisfy (1) --- (4).
\end{enumerate}

(Condition (1) is redundant, as $i \notin S$ and $a_1, \dots, a_m \in S$.) We now claim that any qualifying element $x \neq i$ such that $w(x) < w(a_{j+1})$ must satisfy $\{x, a_{j+1}, \dots, a_m \} \in \mathcal I$. For, suppose $x$ violates some $\mu(B')  = k$, for $B' \subseteq B$. Then this implies that among $\{a_{j+1}, \dots, a_m \}$ there are exactly $k$ elements in $B'$. In particular, $x$ does not qualify for $B' \subseteq B$.

Now consider the set $X \subseteq U$ consisting of all elements $x$ which satisfy 
$$
x \neq i, w(x) < w(a_{j+1}), \{x, a_{j+1}, \dots, a_m \} \in \mathcal I.
$$
As we have seen, $a_j$ is the element of $X \cap S$ of largest weight and $n_j$ is the number of elements of $X$ of greater weight than $a_j$. 

If $|X| \leq n_j$, then the probability that $|N_j| = n_j$ is zero. Otherwise, we can view this as the following process. Suppose we sort the elements of $X$ in order of decreasing weight. Starting with the largest element of $X$, we assign elements to either $S$ or $T$. These assignments to $S$ are independent with probability $1-p$. Then $|N_j| = n_j$ iff we assign the first $n_j$ elements to $T$ (probability $p$) and the $(n_j+1)$th (if it exists) to $S$, which occurs with probability at most $p^{n_j}$.

Hence, conditional on any $a_{j+1}, \dots, a_m$, the probability that $|N_j| = n_j$ is at most $p^{n_j}$.

\end{proof}

\subsection{Probability of selecting an item}
Define the \emph{backward rank} of element $i$ for $B \in \mathcal{F}$, denoted as $\text{brank}(i,B)$, to be the number of elements in $\text{OPT}(B)$ having weight less than $w_i$. Similarly, let $\text{brank}_S(i,B)$ be the number of elements is $\text{OPT}_S(B)$ having weight less than $w_i$. It can be easily seen that $\text{brank}_S(i,B) \geq \text{brank}(i,B)$. Furthermore, if $i \in T$, then $\text{brank}_S(i,B) \geq \text{brank}(i,B) + 1$ (proved in \cite{laminar}). Intuitively, an element $i$ is more likely to be picked by the algorithm if its $\text{brank}(i,B)$ is large.

Now, when element $i \in T$ is considered for inclusion in the solution set, it will be rejected iff there is some $B \in \mathcal{F}(i)$ such that all elements in $OPT_S(B)$ of weight less than $w(i)$ have been evicted already. Let {\sc AllKicked}$(i,B)$ denote this bad event. We can bound the probability of this event as follows.
\begin{lemma}
Suppose $p < 1/2$. Consider any $B \in \mathcal F$ and $i \in OPT$. Now if we define
\begin{align*}
\alpha &= (\frac{p+(1-p) \log (1-p)}{2 (1-p) p^2}) \\
c &= 4 p (1-p)
\end{align*}
Then we have
$$
P({\text{\sc AllKicked}}(i,B)) \leq \frac{\alpha c^{\text{brank}(i,B)+1}}{1-c}
$$

\end{lemma}
\begin{proof}
Fix some $i \in OPT$ and let $\text{brank}(i,B) = d$. All the probabilities we calculate in this proof are conditioned on $i \notin S$; we no longer specify this explicitly to simplify the notation.

Let $\text{OPT}_S(B) = \{a_1, \dots, a_m \}$ sorted so that $w(a_1) < w(a_2) < \dots < w(a_m)$.  Because of the KickNext rule, the item $i$ will go into $SOL(B)$ unless, for some $l \in \{ d+1, \dots, m \}$, there have been at least $l$ items of weight less than $w(a_{l+1})$ added to $\text{SOL}(B)$ before it.

Now consider an element $i' \neq i$. In order for such an $i'$ to have been added to $\text{SOL}(B)$ before $i$, the following events must have occurred:
\begin{enumerate}
\item $i'$ is qualifying for $B$
\item  $i'$ comes before $i$ in the ordering $\pi$
\end{enumerate}

We view the suffix of the permutation $\pi$ corresponding to $T$ as generated by the following process. Each element $x \in T$ chooses $\rho(x)$ uniformly at random from the real interval $[0, 1]$. We then form the suffix of $\pi$ by sorting by $\rho$. Suppose we condition on a fixed value of $r = \rho(i)$. Now consider an element $i' \neq i$. In order for such an $i'$ to have been added to $SOL(B)$ before $i$, the following events
 must have occured:
\begin{enumerate}
\item $i'$ is qualifying for $B$
\item $\rho(i') < r$.
\end{enumerate}

 Let $Q_l$ denote the number of qualifying items other than $i$ with weight $< w(a_{l+1})$ and let $A_l$ denote the number of such items which also have $\rho(i') < i$. We wish to estimate the probability $A_l \geq l$.
 
By Lemma~\ref{Aquallemma}, the random variable $Q_l$ is stochastically dominated by the sum of $l$ independent geometric-$p$ random variables. Given a fixed value for $Q_l$, each such qualifying item $i'$ has a probability $r$ of occuring before $i$. Furthermore, these events are independent (conditional on $r$). Hence the probability $P(A_l \geq l | Q_l = k, i \notin S)$ is at most the probability that a binomial random variable, of $k$ trials and probability $r$, exceeds $l$. In effect, the random variable $A_l$ is formed by conjugating a negative binomial random variable $Q_l$ with a binomial-$r$ distribution. The binomial distribution is a conjugate prior for the negative binomial, hence the distribution of $A_l$ is stochastically dominated by the negative binomial distribution of probability $q = \frac{r p}{1 - p + rp}$.

We now wish to estimate the probability that $A_l \geq l$. For a negative binomial random variable $A'_l$, the event $A'_l \geq l$ is equivalent to the situation that we flip a biased coin for $2 l - 1$ times, where the probability of success is $q$, and the total number of successes is at least than $l$; this is a binomial tail probability. Hence we have
$$
P(A_l \geq l \mid \rho(i) = r) \leq P(\text{Binomial}(2 l -1, q) \geq l-1)
$$

Note that as $p < 1/2$, we have $q < 1/2$ as well. By the Chernoff bound the probability of such a deviation is $\exp(-(2 l - 1) \text{RelEnt}(\frac{l}{2 l - 1} || q))$. Here $\text{RelEnt}$ is the relative entropy function, given by
$$
\text{RelEnt}(x || y) = x \log (x/y) + (1-x) \log(\frac{1-x}{1-y})
$$

We can simplify this as
\begin{align*}
P(A_l \geq l | \rho(i) = r) &\leq \exp(-(2 l - 1) \text{RelEnt}(\frac{l}{2 l - 1} || q)) \\
&= \left(\frac{1-l}{(2 l-1) (q-1)}\right)^{1-l} \left(-\frac{l}{q-2
   l q}\right)^{-l} \\
&\leq \frac{1}{2 - 2q} (4 q (1-q))^l
\end{align*}

Integrating over $r \in [0,1]$ gives
\begin{align*}
P(A_l \geq l) \leq & \int_r \frac{dr}{2 - 2q} (4 q (1-q))^l \\
&\leq (4 p (1-p))^l \int_r \frac{dr}{2 - 2q} \frac{4 q (1-q)}{4 p (1-p)} \\
&\leq (-\frac{p-(p-1) \log (1-p)}{2 (p-1) p^2}) (4 p (1-p))^l \\
&= \alpha c^l 
\end{align*}

We use the union-bound for the event {\sc AllKicked}$(i,B)$:
\begin{align*}
P(\text{\sc AllKicked}(i,B)) &\leq \sum_{l = d+1}^{\infty} P(A_l \geq l \mid i \notin S) \\
&\leq \sum_{l=d+1}^{\infty} \alpha c^l \\
&\leq \frac{\alpha c^{d+1}}{1-c}
\end{align*}
\end{proof}

\subsection{Expected weight of $\text{SOL}$}
We cannot take any arbitrary element of the optimal solution and show that it is selected with a good probability by our matroid secretary algorithm. Instead, we use a similar strategy to the uniform matroid, and examine the set of high-scoring elements \emph{collectively}. We show that most of these elements (but not any particular one of them) are selected high probability.

We contrast our approach with that of \cite{laminar}, which adopted a hybrid proof strategy between fully analyzing the collective behavior of the optimal solution, and analyzing individual elements of the solution. In \cite{laminar}, certain elements in the optimal solution were identified, referred to as ``good'' elements, which were shown to have a high probability of being selected by the secretary algorithm. This type of analysis is inherently not tight. We will instead determine the worst possible arrangement of the optimal solution, and show that it still is selected with high probability.

We use our  upper bound on the probability of the event {\sc AllKicked} to obtain a lower bound on the expected weight of our solution: 
{\allowdisplaybreaks
\begin{align*}
E[w(SOL)] &\geq \sum_{i \in \text{OPT}} w(i) [\text{Probability that }i \in \text{SOL}]\\
&\geq \sum_{i \in \text{OPT}} w(i) \times p \times [\text{Probability that }i \in \text{SOL} | i \notin S]\\
&\geq \sum_{i \in \text{OPT}} w(i) \times p \times [1 - \sum_{B \in \mathcal{F}(i)} P(\text{\sc AllKicked}(i,B))]\\
&\geq p\left[\sum_{i \in \text{OPT}} w(i) - \sum_{i \in \text{OPT}}\sum_{B \in \mathcal{F}(i)} w(i).\dfrac{\alpha c^{1+\text{brank}(i,B)}}{1-c} \right]\\
&\geq p\left[w(\text{OPT}) - \dfrac{\alpha}{1-c}\sum_{i \in \text{OPT}} w(i) \sum_{B \in \mathcal{F}(i)} c^{1+\text{brank}(i,B)} \right]
\end{align*}
}
In order to use this estimate, we need to obtain an upper bound on the sum
$$
\sum_{i \in \text{OPT}} w(i) \sum_{B \in \mathcal{F}(i)} c^{1+\text{brank}(i,B)}
$$

The presence of the weight $w(i)$ complicates things, so as a preliminary we consider the unweighted version of this sum.

Let $\text{OPT}_{\text{large}}^m(B)$ denote the $m$ largest elements in $\text{OPT}(B)$.
\begin{lemma}
Let $B \in \mathcal F$ and let $m \geq 0$ be an integer. Define $g(m,B)$ by
$$
g(m, B) = \sum_{i \in \text{OPT}_{\text{large}}^m(B)} \sum_{B' \in \text{Chain}[M(i), B]} c^{1 + \text{brank}(i,B')}
$$
Suppose $c < 1/2$. Then $$
g(m, B) \leq \frac{2 c}{1-c} | \text{OPT}_{\text{large}}^m (B) |.
$$

\end{lemma}
\begin{proof}
For each integer $i$ define $c_i = c + c^2 + \dots + c^i$, and define $c_{\infty} = \dfrac{c}{1-c}$.

We will need to show a stronger bound, specifically that for all $B \in \mathcal F$ and all $m \geq 0$ we have
$$
g(m, B) \leq 2 c_1 + \dots + 2 c_{m-1} + c_m + c_m c_{k - m}
$$
where $k = \mu(B) \geq m$.

We will show this by induction on the capacity $k$. Note that for a given value of $k$, we are proving the inductive hypothesis simultaneously for all $B \in \mathcal F$ and all possible values of $m$.

We view the laminar family as consisting of levels, corresponding to each possible value for the capacity. When computing $g(m,B)$, we have the contribution at level $k$ itself, as well as the contribution from the lower levels. Let $B_1, \dots, B_j$ be a coarsest $\mathcal F$-partition of $B$ (other than $B$ itself). Let $X = \text{OPT}_{\text{large}}^m(B)$, and let $m_i = |B_i \cap X|$ and $k_i = \mu(B_i)$ for each $i = 1, \dots, j$. For each $i$ we have $X \cap B_i = \text{OPT}_{\text{large}}^{m_i} (B_i)$. By the capacity constraints we must have $m_i \leq k_i < k$ for each $i$.

By laminarity we have
$$
g(m,B) = \sum_{i \in X} c^{1 + \text{brank}(i, B)} + g(m_1, B_1) + \dots + g(m_j, B_j)
$$

The elements of $X$ have maximal bottom-rank in $X$. Hence the term $\sum_{i \in X \cap B} c^{1 + \text{brank}(i, B)} = c^k + \dots c^{k-m+1} = c_m c^{k-m}$.  Each $B_i$ has rank less than $k$ so we apply the inductive hypothesis and obtain
$$
g(m,B) \leq c_m c^{k-m} + \sum_{i=1}^j 2 c_1 + \dots + 2 c_{m_i - 1} + c_{m_i} + c_{m_i} c_{k_i - m_i}
$$

The right-hand side is a convex function $m_1, \dots, m_j$, hence it attains its maximum when these are set to their most extreme possible values. When $m < k$ strictly, we may set $j = 1, m_1 = m, k_1 = k - 1$; when $m = k$, we may set $j = 2, k_1 = k_2 = k-1, m_1 = m-1, m_2 = 1$. In the first case, we obtain
\begin{align*}
g(m,B) &\leq c_m c^{k-m} + \sum_i 2 c_1 + \dots + 2 c_{m_i - 1} + c_{m_i} + c_{m_i} c_{(k-1) - m_i} \\
&\leq c_m c^{k-m} + 2 c_1 + \dots + 2 c_{m - 1} + c_{m} + c_{m} c_{k - 1 -  m} \\
&= 2 c_1 + \dots + 2 c_{m - 1} + c_{m} +  c_{m} ( c_{k - 1 -  m} + c^{k-m} ) \\
&= 2 c_1 + \dots + 2 c_{m - 1} + c_{m} +  c_{m} c_{k - m}
\end{align*}

In the second case, we obtain
\begin{align*}
g(m,B) &\leq c_m c^{k-m} + 2 c_1 + \dots + 2 c_{m-2} + c_{m-1} + c_{m-1} c_{(k-1)-(m-1)} + c_1 + c_1 c_{(k-1)-1} \\
&= c_m + 2 c_1 + \dots + 2 c_{m-2} + c_{m-1} + c_1 + c_1 c_{m-2} \\
&= 2 c_1 + \dots + 2 c_{m-2} + c_{m-1} + c_{m} + c + c c_{m-2} \\
&= 2 c_1 + \dots + 2 c_{m-1} + c_{m} \\
&= 2 c_1 + \dots + 2 c_{m-1} + c_{m} c_{k-m}
\end{align*}
as claimed.
\end{proof}

Next we use this unweighted bound to bound the weighted sum:
\begin{lemma}
If $c < 1/2$ we have
$$
\sum_{i \in \text{OPT}} w(i) \sum_{B \in \mathcal{F}(i)} c^{1+\text{brank}(i,B)}  \leq \frac{2 c}{1-c} w(\text{OPT}) 
$$
\end{lemma}
\begin{proof}
Sort the elements of $\text{OPT}$ by weight so that $w(x_1) > w(x_2) > \dots > w(x_l)$. Define $c_{\infty} = \frac{c}{1-c}$ as above. Then we have
\begin{align*}
&\sum_{i \in \text{OPT}} w(i) \sum_{B \in \mathcal{F}(i)} c^{1+\text{brank}(i,B)} \\
& \qquad = w(x_l) g(l, U) + ( w(x_{l-1}) - w(x_l) ) g(l-1, U) + \dots + (w(x_2) - w(x_1)) g(1, U)  \\
& \qquad \leq w(x_l) 2 l c_{\infty} + ( w(x_{l-1}) - w(x_l) ) 2 (l - 1) c_{\infty} + \dots + (w(x_2) - w(x_1)) 2 c_{\infty}  \\
& \qquad = 2 c_{\infty} ( w(x_l) + w(x_{l-1}) + \dots w(x_1) ) \\
& \qquad = 2 c_{\infty} w(OPT)
\end{align*}
\end{proof}

We consider the contributions to $\text{SOL}$ of the elements of $\text{OPT}$.
\begin{theorem}
\label{Alam-weight-thm}
The expected value of the weight of $\text{SOL}$ is at least a factor
$p (1 - \frac{2 \alpha c}{(1-c)^2} )$ of optimal.
\end{theorem}
\begin{proof}
\begin{align*}
E[w(SOL)] &\geq \sum_{i \in \text{OPT}} w(i) p (1 - \sum_{B \in \mathcal F} P(\text{\sc AllKicked}(i,B))) \\
&\geq p (\sum_{i \in \text{OPT}} w(i) - \sum_{B \in \mathcal F} \sum_{i \in \text{OPT} \cap B} w(i) \frac{\alpha}{1-c} c^{\text{brank}(i,B)+1}) \\
&\geq p (w(\text{OPT}) - \frac{\alpha}{1-c} 2 c_{\infty} w(\text{OPT})) \\
&= w(\text{OPT}) p (1 - \frac{2 \alpha c}{(1-c)^2})
\end{align*}
\end{proof}

\begin{theorem}
The KickNext algorithm achieves a competitive ratio of $0.053$
\end{theorem}
\begin{proof}
Set $p = 0.08$ and apply Theorem~\ref{Alam-weight-thm}.
\end{proof}

\end{document}